\documentstyle[prd,aps,multicol]{revtex}
\newcommand{\nA}{\rlap{/}\!A}

\begin{document}

\title{{\bf Role of Ambiguities and Gauge Invariance 
in the Calculation of the Radiatively Induced
Chern-Simons Shift in Extended Q.E.D.}}
\author{O.A. Battistel* \and G. Dallabona** }
\maketitle

\centerline{* Physics Department}

\centerline{Universidade Federal de Santa Maria}

\centerline{P.O. Box 5093, 97119-900, Santa Maria, RS, Brazil}

\vskip1cm

\centerline{** Physics Department-ICEx}

\centerline{Universidade Federal de Minas Gerais}

\centerline{P.O. Box 702, 30161-970, Belo Horizonte, MG, Brazil}

\begin{abstract}
We investigate the possibility of Lorentz and CPT violations in the photon sector, of the 
Chern-Simons form, be induced by radiative corrections arising from the Lorentz and CPT 
non-invariant fermionic sector of an extended version of QED. By analyzing the modified 
vacuum polarization
tensor, three contributions are considered: two of them can be identified with well known 
amplitudes; the (identical) QED vacuum polarization tensor and the (closely related) $AVV$ 
triangular amplitude. These amplitudes are evaluated in their most general form (to include 
in our discussion automatically the question of ambiguities) on the point of view of a strategy 
to manipulate and calculate divergent amplitudes that can avoid the explicit calculation of 
divergent integrals. Rather than this only general properties are used in intermediary
steps. With this treatment, the results obtained by others authors can be easily recovered and 
we show that, if we choose to impose U(1) gauge invariance maintenance in the pure QED calculated 
amplitudes, to be consistent with the renormalizability, the induced Chern-Simons term assumes 
a nonvanishing ambiguities free value. However if, in addition, we choose to get an answer 
consistent with renormalizability by anomaly cancellation of the Standard Model a vanishing 
value can be obtained, in accordance with what was previously conjectured by other authors.
\end{abstract}

\noindent
PACS 11.30.Cp
\vskip0.5cm

\section{Introduction}

The essential ingredient for the construction of Quantum Field Theories are
the symmetries. Undoubtedly, the most important of them all is the Lorentz
invariance, which represents our conception in respect of space-time. This
symmetry has a privileged status when compared to those denominated as internal
symmetries. It is fair to say that is lies at the root of practically all
success in describing the available phenomenology. In other hand, CPT
symmetry has played a crucial hole in the construction of our present
theoretical knowledge about the phenomenology of fundamental particle
interaction. In particular, for the construction of the standard model
Lagrangian, Lorentz and CPT invariance are required in addition to
invariance under $SU(3)\otimes SU(2)\otimes U(1)$ gauge transformations 
\cite{1}. 
From
the point of view of this symmetry content, all possible interactions involving
an arbitrary number of fundamental fields, simultaneously invariant under
Lorentz, CPT and gauge transformations, are in principle realizations of the
model. Other two restrictions reduce drastically the set of interactions
suggested by the adopted symmetry content and fixes the number of fundamental 
quantum fields. The first one is the renormalizability by power counting
that eliminates all invariant interaction with canonical dimension $d\succ 4$.
The second one is the renormalizability by anomaly cancellation, which forces us
to find a number of fermion fields to the theory in such a way the
violations in symmetry relations introduced by the anomalies cancel each
other to maintain the renormalizability of the model \cite{2}. It is very important
to emphasize that these two crucial restrictions above cited do not resides
in the symmetry content but are deeply related to the perturbative solution
of the theory and therefore to practical reasons. They result directly from
our limitations in the treatment of divergences emerging in the perturbative
approach. In this context the success of the resulting theory in the
accurate description of the corresponding phenomenology is a direct
consequence of the validity of the gauge symmetries Lorentz and CPT on the
interaction of the fundamental fields conditioned to the restrictions
that lead to a renormalizable theory. In the case of Lorentz and CPT
symmetries, to the present all experimental indications point to absolute
symmetries \cite{3}. However the technological evolution nowadays allows us to
test the limit of validity of these particular symmetries to a crescent high
degree of precision. Having this in mind the implication of Lorentz and CPT
violations have been receiving a lot of attention, specially after the work of
Colladay and Kosteleck\'{y} \cite{4} where a conceptual framework and a procedure for
treating spontaneous CPT and Lorentz violations is developed (maintaining
gauge structure and renormalizability). Within this framework a CPT violating
extension of the minimal Standard Model is presented and some
phenomenological consequences established. In a more recent work Colladay and
Kosteleck\'{y} \cite{5} have continued their theoretical investigations and presented a
full Lorentz-violating extension for the Standard Model including CPT-even
Lorentz-breaking terms not explicitly presented in their first work. In
particular an extended version of the quantum electrodynamics is extracted 
from the above cited model in such a way that Lorentz and CPT
violations are included. As a consequence the usual properties of the fermion
and the photon are modified and many measurable implications are considered
and discussed. Additional investigations in the subject of breaking the Lorentz
invariance have been performed by Colleman and Glashow \cite{6} in a work
based on the more general theory of Colladay and Kosteleck\'{y} \cite{4}, 
\cite{5}, but
restricted to the special case of rotational and CPT invariance. Tiny Lorentz
breaking terms are included in the Lagrangian of the Standard Model in such
a way that $SU(3)\otimes SU(2)\otimes U(1)$ gauge invariance, renormalizability 
by power counting and by anomaly cancellation are maintained. A lack of
possible experimental tests of the Lorentz breaking are considered in detail
in the high energy regime. 

In this very general discussion on the
possibilities of breaking Lorentz and CPT invariance one problem received
special attention; the possibility of to induce these breaking by 
radiative corrections in an extended QED, a sector of the extended version of
the Standard Model. The Lagrangian of the extended QED, given in the Colladay and
Kosteleck\'{y} \cite{5} investigations, is composed by the usual QED Lagrangian 
adding
three other contributions coming from the breaking possibilities:
\[
L^{SB} = - a_{\mu}\bar{\Psi}\gamma^{\mu}\Psi - b_{\mu}\bar{\Psi}\gamma_{5}\gamma^{\mu}\Psi
+ \frac{1}{2}k^{\alpha}\epsilon_{\alpha \lambda \mu \nu} A^{\lambda} F^{\mu \nu}
\]
where $a_\mu$ and $b_\mu $ are constant (real) prescribed four-vectors, the
coupling $k^\alpha $ is real and has dimensions of mass and $\gamma _5$ is the
usual Dirac Hermitian matrix related to $\epsilon _{\alpha \lambda \mu \nu
}$ by $tr{\gamma_5 \gamma_\alpha  \gamma_\beta  \gamma_\mu \gamma_\nu} = 
4i\epsilon _{\alpha \beta \mu \nu }$. An important point about the
breaking term of the photon sector in the above Lagrangian is the behavior
under potential gauge transformation $(A_\mu \rightarrow A_\mu +\partial_\mu
\Lambda )$ \cite{7}. It change by a total derivative leading to an invariant 
action and to the same equations of motion. This behavior characterize the Chern-Simons
form \cite{8}. There are theoretical and experimental aspects related to the
phenomenology predicted by the modified theory \cite{9}, \cite{4}-\cite{6}, 
\cite{10}. All present analysis seems
to point to a zero value for the $k_\mu $ coupling. In searching from
Lorentz and CPT effects an immediate question emerges in light of vanishing
of the $k_\mu $ in the tree level: is whether the radiative correction coming
from another sector of the theory can induce contributions of the
Chern-Simons type. This contribution would represent a correction to the
photon propagator \cite{5}. From the point of view of the perturbative 
diagrammatic
(Feynman rules) expansion the diagrams corresponding to the modified theory
are closely related to the ones of the symmetric theory. They are obtained by
insertions on propagators and vertices in the topologies of symmetric theory. The
lowest order diagram that represents the correction to the photon propagator is
the usual QED vacuum polarization tensor with an insertion or on the internal
charged fermion propagator or at the vertices of the diagram. In different
words, we need to evaluate the usual one-loop vacuum polarization amplitude
but changing the free fermion propagator $S(k)$, obeying the Dirac equation,
to a propagator coming from the addition of the $b_\mu $ coupling to the
usual QED Lagrangian;
\begin{equation}
G(k)=\frac{i}{\not{k}-m-\not{b}\gamma_{5}}.
\end{equation}
Explicitly, we are lead to evaluate the amplitude: 
\[
\Pi ^{\mu \nu }(p)=\int \frac{d^4k}{(2\pi )^4}tr\left\{ \gamma ^\mu G(k)\gamma
^\nu G(k+p)\right\}. 
\]
The evaluation of this radiative correction has been performed by different
authors \cite{5}, \cite{7}, \cite{11}, \cite{12}, \cite{13} emphasizing 
many aspects involved in the calculations. A common
conclusion emerges in all of these investigations; the result obtained is not
free from ambiguities. The main reason for this is due to the fact that the
resulting amplitude $\Pi ^{\mu \nu }(p)$ is a divergent object, and in this
way, plaged by the very well known problems common to the perturbative
calculations in quantum field theories. To evaluate a divergent amplitude
the first step we need is to specify a mathematical procedure in which the
calculations are turned to be possible. Usually this means to adopt a
regularization technique or equivalent philosophy. The choice of such a scheme
is guided by the consistency of the procedure in preserving underlying
features of the specific theory or general aspects of quantum field theory.
There are two dramatic problems involved in the desired consistency: 
maintenance of symmetry relation and the avoidance of ambiguities related to
possible arbitrary choices in the internal momentum routing of the divergent
amplitudes. The first aspect is crucial to get a renormalizable theory and
second is crucial to establish the power prediction of a quantum field theory,
both aspects related to the perturbative solutions. In this context,
undoubtedly, the most important technique, specially for the development of
gauge quantum field theory, is the Dimensional Regularization \cite{14}. Unfortunately
this is not a general procedure because, among others, the implementation of
the $\gamma _5$ matrix is not well defined \cite{15} and,  consequently, when this
object is present we need to recourse to another procedure. The most famous
problem involving this situation is that of triangle anomalies where, when the
perturbative evaluation is made in the context of four-dimensional procedure, 
ambiguities are explicitly isolated \cite{16}.

This briefly discussion about these general aspects involved in the evaluation
of divergent amplitudes is needed to establish a point of view to our analysis
as will became clear in what follows. As have been pointed out by Colladay and
Kosteleck\'{y} \cite{5} the two point diagram $\Pi _{\mu \nu }(p)$, written above with
the inclusion of the breaking correction in the propagator, can be viewed in
the context of the more fundamental theory from which QED extended has been
extracted, as a mathematical structure identical to a one-loop three
point diagram coupling to a two photon lines and one axial vector. We can
look to the two point diagram as a kinematical limit of the corresponding
three point diagram in which the momentum associated to axial-vector leg is
zero. With this identification we are lead to evaluate in our calculation
mathematical structures identical to two well known amplitudes: the usual
QED vacuum polarization tensor and the $AVV$ amplitude related to chiral
anomalies and to pion decay phenomenology \cite{7} .

These statement can be put in a very clear way if we write the exact 
propagators $G(k)$, given in the expression (1), as in the work of Jackiw 
and Kosteleck\'{y} \cite{7}, in the form
\[
G(l)=S(l)+G_{b}(l)
\]
where:
\[
G_{b}(l)=\frac{i}{\not{k}-m-\not{b}\gamma_{5}} \not{b}\gamma_{5}S(l)
\]
In this way the amplitude $\Pi_{\mu\nu}(p)$ can be split in three terms
\begin{equation}
\Pi^{\mu \nu}=\Pi_{0}^{\mu \nu}+\Pi_{b}^{\mu \nu}+\Pi_{bb}^{\mu \nu}
\end{equation}
where $ \Pi_{0}^{\mu \nu}$ is the usual QED vacuum polarization tensor  and
$\Pi_{b}^{\mu \nu}$  is given by
\[
\Pi_{b}^{\mu \nu}(p)= \int\frac{d^4k}{(2\pi)^4}tr\left\{ \gamma^\mu S(l) \gamma^\nu G_b(l+p)
+ \gamma^\mu G_b(l+p) \gamma^\nu S(l) \right\}
\]  
where the identification with the mathematical structure of the $AVV$
amplitude, as described above, became now clear. The possible contribution to
the induced Chern-Simons term linear in $b_\mu $ $(b_\mu \propto k_\mu )$ come
from $\Pi_{b} ^{\mu \nu }(p)$. The results of the investigations 
\cite{5}, \cite{6}, \cite{7}, \cite{11}, \cite{12}, \cite{13} 
have been generating two types of controversy: the first it refers to the existence of
such CPT and Lorentz violation correction; the other as what is the value
of the correction. The first aspect is related to general matters. Colleman and
Glashow \cite{5} argued that only a vanishing value of $k_\mu $ is allowed if one
demands that the axial current $j_5^\mu (x)=\bar{\Psi}(x)\gamma ^\mu \gamma
^5\Psi (x)$ should maintain gauge invariance in a Quantum Field Theory for
any value of the momenta, or equivalently, at any position $x$. As a
consequence of such hypothesis the value of $k_\mu $ should be unambiguous
vanishing to first order in $b_\mu $ for any CPT-odd, gauge invariant
interaction. On the other hand, Jackiw and Kosteleck\'{y} \cite{7} argued that a
weaker condition could be assumed; in the case when there is no coupling of
the axial current with other fields then the physical gauge invariance would
be maintained at zero momentum. Put in a different way, the action can be
gauge invariant although the Lagrangian density is not. From this point of
view $k_\mu $ does not needs to vanish. In this case, what would be its
value? This whole question is intimately connected with the fact that, in
order to evaluate the correction through perturbative techniques, one needs
to deal with divergent integrals. As we have pointed above it is 
therefore essential, at some point,
the definition of a regularization strategy in order to perform the
calculation. The final result is clearly subject to the usual problems
involved in the treatment of divergent integrals such as ambiguities, for
example. For this reason different values for $k_\mu $ have been suggested
in the literature. In their paper, Jackiw and Kosteleck\'{y} \cite{7} 
explicitly
calculate this contribution and find a nonvanishing value. Their calculation
is strongly based in the ambiguous character of the mathematical structures
involved, when explicitly taking into account surface terms associated to
shifts effected in the arbitrary internal momentum routing of the loops. At
this point we can say that in all works where the radiatively Chern-Simons
term was calculated no unique answer can be given. In a certain way this is
not a surprising fact because we are dealing with the typical mathematical
indefinitions of the divergences of perturbative 
solutions of quantum field theories. So, the answer to this question, as well
as to any one involving a particular divergent amplitude, cannot be given in
a individually way. This means that any regularization scheme or equivalent
procedure used to evaluate a particular amplitude needs to be tested in a
more general context to prove its consistency. In particular in the QED
extended theory the U(1) gauge symmetry is maintained. This implies that the
Ward identities associated to vector current conservation (gauge invariance)
need to be present in all calculated amplitudes. If this is not the case 
the renormalizability of the usual QED would be
spoiled by the eventual procedure applied and in consequence the
renormalizability of the general theory, the Standard Model, is lost too.
Another very important fact related to the mathematical structures involved
in the discussed problem, that is crucial for the renormalizability of the
full model, is the anomaly cancellation, once we have verified that the
linear $b_\mu $ term is closely related to triangular anomaly diagrams. So,
we believe that to answer the question related to what is the value for the
radiatively induced Chern-Simons term the most general calculation must be
given. However only after the consistency of this particular calculation are put in
accordance with the more general context a value can be extracted. A
negative affirmation is included in this statement: if a procedure do not
furnishes a consistent result for a particular amplitude of the theory then the
results produced by this procedure to any other amplitude cannot be taken
seriously for any purposes. 

The purpose of this contribution is to investigate in a more general context of
perturbative calculations
what is the possible role played by the ambiguities when the
general aspects related to renormalizability are considered. To evaluate the
amplitude we will use a strategy to calculate and manipulate divergent integrals 
in which no divergent integrals are, in fact, calculated. In the result 
produced by our technique it will be possible
to map the results produced by others authors using different schemes. 

The specific implication of our argument to the discussed problem is that we need to evaluate 
the term $\Pi _0^{\mu \nu }$ and $\Pi _b^{\mu \nu }$ in the expression (2) in a way that the 
results produced in both cases are required to be consistent simultaneously. Our main guide 
in this investigation is the maintenance of the U(1) gauge symmetry in the usual QED theory. 
In order to define a consistent strategy to handle the divergent amplitudes we first treat the
vacuum polarization tensor $\Pi _0^{\mu \nu }$ in such a way that any specific assumption 
about a regulator is avoided. In intermediary steps we only make use of very general 
properties of a such eventual regulator. The divergent content of the amplitudes is 
separated from the finite one by the use of a identity at the level of the integrand. 
In the final form so obtained, no dependence with external momenta are present in a set 
of basic divergent objects that we will define. In respect to finite terms, they are in 
this way, free from effects of regularization. In our final results those of Jackiw and 
Kosteleck\'{y} \cite{7} will be recovered in detail. After all calculations have been 
performed we show how the choice of demanding of gauge invariance for the usual vacuum 
polarization tensor will lead us to a nonzero value ambiguity free for the Chern-Simons 
term, but if in addition we choose to require consistency with the anomaly cancellation 
a vanishing value for the Chern-Simons radiatively induced contribution can be obtained.

\section{The QED Vacuum Polarization Tensor with Arbitrary Internal Momentum Routing}

Let us now evaluate the well known Q.E.D. vacuum polarization tensor to one 
loop order from the point of view of our proposed strategy \cite{17}. We perform 
the calculations allowing complete arbitrariness in the internal momentum routing. 
Explicitly we have
\begin{equation}
\left( \Pi _0\right) _{\mu \nu }=\int \frac{d^4k}{\left( 2\pi \right) ^4}
tr\left\{ \gamma _\mu S(k+k_1)\gamma _\nu S(k+k_2)\right\} ,
\end{equation}
where $S(k)$ is a usual $1/2$ spin free fermion propagator, carrying momentum $k$. 
Of course if an ambiguity free result is required only the combination $(k_1-k_2)$ 
should be allowed. Next, we can put $\Pi _0^{\mu \nu }$ in a more convenient form 
for our purposes. Explicitly: 
\begin{eqnarray}
\left( \Pi _{0}\right) _{\mu \nu } &=&4\left\{
\int_{\Lambda }\frac{d^{4}k}{(2\pi )^{4}}\frac{2k_{\mu }k_{\nu }}{%
[(k+k_{1})^{2}-m^{2}][(k+k_{2})^{2}-m^{2}]}\right.  \\
&&+(k_{1}+k_{2})_{\nu }\int_{\Lambda }\frac{d^{4}k}{(2\pi )^{4}}\frac{k_{\mu
}}{[(k+k_{1})^{2}-m^{2}][(k+k_{2})^{2}-m^{2}]}  \nonumber \\
&&+(k_{2}+k_{1})_{\mu }\int_{\Lambda }\frac{d^{4}k}{(2\pi )^{4}}\frac{k_{\nu
}}{[(k+k_{1})^{2}-m^{2}][(k+k_{2})^{2}-m^{2}]}  \nonumber \\
&&+\left. \left( k_{2\mu }k_{1\nu }+k_{1\mu }k_{2\nu }\right) \int_{\Lambda }%
\frac{d^{4}k}{(2\pi )^{4}}\frac{1}{[(k+k_{1})^{2}-m^{2}][(k+k_{2})^{2}-m^{2}]%
}\right\}   \nonumber \\
&&-8g_{\mu \nu }\left\{ \int_{\Lambda }\frac{d^{4}k}{(2\pi )^{4}}\frac{1}{%
[(k+k_{1})^{2}-m^{2}]}+\int_{\Lambda }\frac{d^{4}k}{(2\pi )^{4}}\frac{1}{%
[(k+k_{2})^{2}-m^{2}]}\right.   \nonumber \\
&&\left. -(k_{1}-k_{2})^{2}\int_{\Lambda }\frac{d^{4}k}{(2\pi )^{4}}\frac{1}{%
[(k+k_{1})^{2}-m^{2}][(k+k_{2})^{2}-m^{2}]}\right\} .  \nonumber
\end{eqnarray}
To obtain the above expression we have evaluated the traces over the Dirac matrices 
and used the following identity: 
\begin{eqnarray}
(k+k_i)\cdot (k+k_j)=\frac 12\left[ (k+k_i)^2-m^2\right] +\frac 12\left[
(k+k_j)^2-m^2\right] -\frac 12\left[ (k_i-k_j)^2-2m^2\right] .
\end{eqnarray}

In the expression (4) we can identify a set of divergent integrals with degree of divergence 
that includes linear and quadratic ones. In this way, special attention needs to be given to 
the question of ambiguities related to possible shifts in the loop momentum. At this point the 
usual procedure is to adopt some regularization philosophy. To avoid specific choices from now 
on we will consider all divergent integrals to be regulated in 4-D by some implicit function, 
such that
\begin{equation}
\int \frac{d^{4}k}{\left( 2\pi \right) ^{4}}f(k)\rightarrow \int \frac{d^{4}k%
}{\left( 2\pi \right) ^{4}}f(k)\left\{ \lim_{\Lambda _{i}^{2}\rightarrow
\infty }G_{\Lambda _{i}}\left( k^{2},\Lambda _{i}^{2}\right)
\right\}=\int_{\Lambda }\frac{ d^{4}k}{\left( 2\pi \right) ^{4}}f(k).
\end{equation}
Here $\Lambda_i^{\prime}s $ are parameters of a distribution $ G(k^2,\Lambda^2_i) $ whose 
asymptotic behavior in $k$ renders the integral finite. Besides this, we require the existence 
of a connection limit
\begin{equation}
\lim_{\Lambda _i^2\rightarrow \infty }G_{\Lambda _i}\left( k^2,\Lambda
_i^2\right) =1.
\end{equation}
This condition, in particular, guarantees that the finite integrals value will not be modified.
We can make use of this property to extract the part of the amplitude that carries the external
momentum dependence without any contamination by regularization. Having this in mind, after
adopting the implicit presence of a generic distribution, we treat all integrals by using 
identities at level of the integrand in such a way as to obtain a mathematical structure where 
the dependence on external momenta are contained only in finite integrals 
\footnote{The philosophy is somewhat close in spirit to the BPHZ procedure \cite{18} 
where a Taylor expansion is made around the value of the physical momentum $p=0$.}. 
In its most general form this identity reads
\begin{equation}
\frac 1{[(k+k_i)^2-m^2]}=\frac 1{(k^2-m^2)}+\sum_{j=1}^N\frac{\left(
-1\right) ^j\left( k_i^2+2k_i\cdot k\right) ^j}{\left( k^2-m^2\right) ^{j+1}}
+\frac{\left( -1\right) ^{N+1}\left( k_i^2+2k_i\cdot k\right) ^{N+1}}{\left(
k^2-m^2\right) ^{N+1}\left[ \left( k+k_i\right) ^2-m^2\right] },
\end{equation}
where $k_i$ is an arbitrary momentum. The value of $N$ should be chosen so that the last 
term in the above expression now corresponds to a finite integral, and then, by virtue 
of (7), the integration can be performed without restriction (the integrals so handled 
can be found in appendix B). With respect to divergent parts, no additional supposition 
is made and all terms are written as combinations of five basic objects \cite{17}, 
\cite{19}: 
\begin{eqnarray}
\bullet \Box _{\alpha \beta \mu \nu } &=&\int_\Lambda \frac{d^4k}{\left( 2\pi
\right) ^4}\frac{24k_\mu k_\nu k_\alpha k_\beta }{\left( k^2-m^2\right) ^4}%
-g_{\alpha \beta }\int_\Lambda \frac{d^4k}{\left( 2\pi \right) ^4}\frac{%
4k_\mu k_\nu }{\left( k^2-m^2\right) ^3} \\
&&-g_{\alpha \nu }\int_\Lambda \frac{d^4k}{\left( 2\pi \right) ^4}\frac{%
4k_\beta k_\mu }{\left( k^2-m^2\right) ^3}-g_{\alpha \mu }\int_\Lambda \frac{%
d^4k}{\left( 2\pi \right) ^4}\frac{4k_\beta k_\nu }{\left( k^2-m^2\right) ^3}
\nonumber \\
\bullet \Delta _{\mu \nu } &=&\int_\Lambda \frac{d^4k}{\left( 2\pi \right) ^4}\frac{%
4k_\mu k_\nu }{\left( k^2-m^2\right) ^3}-\int_\Lambda \frac{d^4k}{\left(
2\pi \right) ^4}\frac{g_{\mu \nu }}{\left( k^2-m^2\right) ^2} \\
\bullet \nabla _{\mu \nu } &=&\int_\Lambda \frac{d^4k}{\left( 2\pi \right) ^4}\frac{%
2k_\nu k_\mu }{\left( k^2-m^2\right) ^2}-\int_\Lambda \frac{d^4k}{\left(
2\pi \right) ^4}\frac{g_{\mu \nu }}{\left( k^2-m^2\right) }
\end{eqnarray}
\begin{eqnarray}
\bullet I_{log}(m^2) &=&\int_\Lambda \frac{d^4k}{\left( 2\pi \right) ^4}\frac
1{\left( k^2-m^2\right) ^2}   \\
\bullet I_{quad}(m^2) &=&\int_\Lambda \frac{d^4k}{\left( 2\pi \right) ^4}\frac
1{\left( k^2-m^2\right) }.
\end{eqnarray}
This way it is possible to make contact with the 
corresponding results with other traditional techniques if the above objects are evaluated 
in the adopted scheme. The main advantage of this strategy is to manipulate each divergent 
integral without explicit calculations of a divergent integral. This decomposition 
is 
sufficient for the required analysis, and the same expression for a given integral is used in 
all places where it occurs. The price we pay is that the finite integrals at first sight 
seem to be more complicated although they can be readily organized and calculated using 
standard techniques. After this essential discussion we turn our attention back to the 
calculation of $\Pi _0^{\mu \nu }$. Using the results for the integrals obtained according 
to this procedure (appendix B) we can write: 
\begin{eqnarray}
\left( \Pi _0\right) _{\mu \nu }&= &
\frac{4}{3}[(k_1-k_2)^2g_{\mu\nu}-(k_1-k_2)_\mu (k_1-k_2)_\nu ]\times 
\nonumber \\
& &\times\left\{ 
[I_{log}(m^2)]-\left(\frac{i}
{(4\pi )^2}\right)\left[\frac{1}{3}+\frac{(2m^2+(k_1-k_2)^2)}{(k_1-k_2)^2}
Z_0(m^2,m^2,(k_1-k_2)^2)
\right]\right\}\nonumber \\
& &+A_{\mu\nu},
\end{eqnarray}
where
\begin{eqnarray}
A_{\mu\nu}
&= &+4[\nabla_{\mu\nu}]+(k_1-k_2)_\alpha (k_1-k_2)_\beta\left[
\frac{1}{3}\Box_{\alpha\beta\mu\nu}
+\frac{1}{3}\triangle_{\mu\beta}g_{\alpha\nu}
+g_{\alpha\mu}\triangle_{\beta\nu}
-g_{\mu\nu}\triangle_{\alpha\beta}
-\frac{2}{3}g_{\alpha\beta}\triangle_{\mu\nu}\right]
\nonumber \\
& &+\left[(k_1-k_2)_\alpha (k_1+k_2)_\beta - (k_1+k_2)_\alpha 
(k_1-k_2)_\beta \right] \left[\frac{1}{3}\Box_{\alpha\beta\mu\nu}+
\frac{1}{3}\triangle_{\mu\beta}g_{\nu\alpha}+\frac{1}{3}\triangle_{\beta\nu}
g_{\alpha\mu}\right]\nonumber \\
& &+(k_1+k_2)_\alpha (k_1+k_2)_\beta \left[\Box_{\alpha\beta\mu\nu}-
\triangle_{\nu\alpha}g_{\mu\beta}-\triangle_{\beta\nu}
g_{\alpha\mu} -3\triangle_{\alpha\beta}g_{\mu\nu}\right].
\end{eqnarray} 

The definition of the finite function $Z_k$ can be found in appendix A. In the above result
it becomes clear that the presence of the objects $\Box ,\nabla$ and $\Delta $, which are 
differences between divergent integrals with the same degree of divergence, renders the Q.E.D. 
vacuum polarization tensor ambiguous and break Ward identities 
$(k_1-k_2)^\mu \left( \Pi _0\right) _{\mu \nu }=
(k_1-k_2)^\nu \left( \Pi _0\right) _{\mu \nu }=0$, 
which represent violation of U(1) gauge invariance. It is important to note that there 
are three different kinds of violating terms. One of them involves an ambiguous combination 
of the momenta $k_1$ and $k_2$. Another one where only
their differences appear (external momenta) and a third where no dependence on such momenta 
are present. In view of this, it is easy to check that it is not possible to restore gauge 
invariance by choosing values for $k_1$ and $k_2$ (arbitrary) momenta. The ONLY possibility we 
have is the requirement that the value assumed by the objects $\Box ,\nabla$ and  $\Delta $ 
should be simultaneously zero. After assuming these ``consistency conditions'' the remaining 
result for $\Pi _0^{\mu \nu }$ can be immediately identified with the corresponding one 
produced by Dimensional Regularization (D.R.) technique (after expressing $I_{log}(m^2)$ 
in this specific mathematical language). This natural mapping is a consequence of the fact 
that our three differences $\Box ,\nabla$ and $\Delta $ vanish in D.R. For a more detailed 
discussion on the ambiguities and symmetry violations in this framework we 
report the reader to references \cite{17}, \cite{19}, \cite{21}.

We can summarize what we have learned in our analysis of this particular amplitude 
(and many others performed elsewhere) by the sentence: There is no chance of consistency 
in a perturbative calculation without requiring what we call the consistency conditions. 
Ambiguities associated with arbitrary choices of internal momenta and gauge invariance 
constitutes only two aspects that can be associated to them \cite{17}, \cite{19}. 
Having this in mind in what follows we calculate the second term in the expression (2), 
using the same prescriptions above.

\section{Explicit Calculation of the Axial-Vector-Vector Triangle}

The first order contribution to the Chern-Simons term comes from the $\Pi_b^{\mu \nu }$ 
amplitude, the second term in expression (2). It is explicitly given by \cite{7} 
\begin{eqnarray}
\Pi _b^{\mu \nu } &=&\int \frac{d^4k}{\left( 2\pi \right) ^4}tr\left\{
\gamma ^\mu S(k)\gamma ^\nu G_b(k+p)\right.  \\
&&\;\;\;\;\;\;\;\;\;\;\;\;\;\;\;\;\;+\left. \gamma ^\mu G_b(k)\gamma ^\nu
S(k+p)\right\},   \nonumber
\end{eqnarray}
where 
\begin{equation}
G_b(k)=\frac 1{\not{k}-m-\not{b}\gamma _5}\not{b}\gamma _5S(k).
\end{equation}
To evaluate $\Pi _b^{\mu \nu }$ to lowest order in $b$, we simply replace the above expression by 
\begin{equation}
G_b(k)=-iS(k)\not{b}\gamma _5S(k).
\end{equation}
Now, the corresponding expression to $\Pi _b^{\mu \nu }$ may be written as: 
\begin{equation}
\Pi _b^{\mu \nu }(p)\simeq b_\lambda \Pi ^{\mu \nu \lambda }(p),
\end{equation}
where 
\begin{eqnarray}
\Pi ^{\mu \nu \lambda }(p) &=&(-i)\int \frac{d^4k}{(2\pi )^4}
tr\left\{\gamma ^\mu S(k)\gamma ^\nu S(k+p)\gamma ^\lambda \gamma _5S(k+p)\right. + \\
&&\;\;\;\;\;\;\;\;\;\;\;\;\;\;\;\;\;\;\;\;\;\;\;\;\;+\left. \gamma ^\mu S(k)\gamma
^\lambda \gamma _5S(k)\gamma ^\nu S(k+p)\right\}. \nonumber
\end{eqnarray}
These two terms can be identified as a particular kinematic situations of the 
Axial-Vector-Vector triangular amplitude \cite{5}, \cite{7}, related by 
(the anomalous) Ward identity to 
pion decay phenomenology \cite{17}. In virtue of this we will consider the most general 
calculation of the $AVV$ amplitude. Only at the end of the calculation we will return 
to the above specific situations. We believe that proceeding this way our conclusions 
may be more general and transparent. We start by giving the definition: 
\begin{equation}
T_{\lambda \mu \nu }^{AVV}=-\int \frac{d^4k}{\left( 2\pi \right) ^4}tr\left\{
\gamma _\mu \left[ (\not{k}+\not{k}_1)- m\right] ^{-1}\gamma _\nu \left[ (\not
{k}+\not{k}_2)- m\right] ^{-1}\gamma _\lambda \gamma _5\left[ (\not{k}+\not
{k}_3)- m\right] ^{-1}\right\}, 
\end{equation}
where $k_1,k_2$ and $k_3$ stand for arbitrary choices of internal lines momenta. 
They are related to external ones by their differences. From now on we will follow 
strictly the same steps used in the preceding section. After taking the Dirac 
traces 
and using the identity (5) we organize the expression in the following 
and convenient form 
\begin{equation}
T_{\lambda \mu \nu }^{AVV}=-4i\left\{ -F_{\lambda \mu \nu }+N_{\lambda \mu
\nu }+M_{\lambda \mu \nu }+P_{\lambda \mu \nu }\right\} ,
\end{equation}
where we have introduced the definitions 
\begin{eqnarray}
\bullet P_{\lambda \mu \nu } &=&g_{\mu \nu }\varepsilon _{\alpha \beta
\lambda \xi }\int_\Lambda \frac{d^4k}{(2\pi )^4}\frac{(k+k_1)_\alpha
(k+k_2)_\beta (k+k_3)_\xi }{[(k+k_1)^2-m^2][(k+k_2)^2-m^2][(k+k_3)^2-m^2]} \\
\bullet F_{\lambda \mu \nu } &=&\int_\Lambda \frac{d^4k}{\left( 2\pi \right)
^4}\{\varepsilon _{\nu \beta \lambda \xi }\left( k+k_1\right) _\mu \left(
k+k_2\right) _\beta \left( k+k_3\right) _\xi  \\
&&\;\;\;\;\;\;\;\;\;\;\;\;+\varepsilon _{\mu \beta \lambda \xi }\left(
k+k_1\right) _\nu \left( k+k_2\right) _\beta \left( k+k_3\right) _\xi  
\nonumber \\
&&\;\;\;\;\;\;\;\;\;\;\;\;+\varepsilon _{\mu \alpha \nu \beta }\left(
k+k_1\right) _\alpha \left( k+k_2\right) _\beta \left( k+k_3\right) _\lambda 
\nonumber \\
&&\;\;\;\;\;\;\;\;\;\;\;\;+\left. \varepsilon _{\mu \alpha \nu \xi }\left(
k+k_1\right) _\alpha \left( k+k_3\right) _\xi \left( k+k_2\right) _\lambda
\right\} \times   \nonumber \\
&&\;\;\;\;\;\;\;\;\;\;\;\times \left\{ \frac 1{\left[ \left( k+k_1\right)
^2-m^2\right] \left[ \left( k+k_2\right) ^2-m^2\right] \left[ \left(
k+k_3\right) ^2-m^2\right] }\right\}   \nonumber \\
\bullet N_{\lambda \mu \nu } &=&\frac{\varepsilon _{\mu \alpha \nu \lambda }}%
2\left\{ \int_\Lambda \frac{d^4k}{\left( 2\pi \right) ^4}\frac{
(k+k_1)_\alpha }{\left[ (k+k_2)^2-m^2\right] \left[ (k+k_1)^2-m^2\right] }
\right.  \\
&&\;\;\;\;\;\;\;+\int_\Lambda \frac{d^4k}{\left( 2\pi \right) ^4}\frac{%
(k+k_1)_\alpha }{\left[ (k+k_1)^2-m^2\right] \left[ (k+k_3)^2-m^2\right] } 
\nonumber \\
&&\;\;\;\;\;\;\ +[2m^2-(k_2-k_3)^2]\times   \nonumber \\
&&\;\;\;\;\;\;\;\;\times \left. \int \frac{d^4k}{\left( 2\pi \right) ^4}
\frac{(k+k_1)_\alpha }{\left[ (k+k_1)^2-m^2\right] \left[
(k+k_2)^2-m^2\right] \left[ (k+k_3)^2-m^2\right] }\right\}   \nonumber \\
\bullet M_{\lambda \mu \nu } &=& m^2\varepsilon _{\mu \nu \alpha \lambda
}\int \frac{d^4k}{\left( 2\pi \right) ^4}\frac{\left\{ \left( k+k_2\right)
_\alpha -\left( k+k_1\right) _\alpha +\left( k+k_3\right) _\alpha \right\} }{
\left[ \left( k+k_1\right) ^2-m^2\right] \left[ \left( k+k_2\right)
^2-m^2\right] \left[ \left( k+k_3\right) ^2-m^2\right] }.
\end{eqnarray}
Looking at the above expression we can identify a set of Feynman integrals some of them 
are divergent. The most severe degree of divergence in this case is the linear one, which 
is contained in $N_{\lambda \mu \nu }$, in the two point function structures. We use here the 
same results adopted to treat the $\Pi _0^{\mu \nu }$ amplitude. After the evaluation of those
integrals corresponding to three point function structures (appendix B) we arrive at the 
followings forms: 
\begin{eqnarray}
\bullet P_{\lambda \mu \nu } &=&0 \\
\bullet M_{\lambda \mu \nu } &=&-\left( \frac{i}{(4\pi )^{2}}\right)
\varepsilon _{\mu \alpha \nu \lambda }m^{2}\left\{ \left( k_{2}-k_{1}\right)
_{\alpha }\left( \xi _{00}-\xi _{01}\right) +\left( k_{3}-k_{1}\right)
_{\alpha }\left( \xi _{00}-\xi _{10}\right) \right\} \\
\bullet N_{\lambda \mu \nu } &=&\frac{\varepsilon _{\mu \alpha \nu \lambda }%
}{4}\left( k_{1}-k_{2}\right) _{\alpha }\left\{ I_{\log }\left( m^{2}\right)
-\left( \frac{i}{\left( 4\pi \right) ^{2}}\right) Z_{0}\left( \left(
k_{1}-k_{2}\right) ^{2};m^{2}\right) \right. \\
&&\;\;\;\;\;\;\;\;\;\;\;\;\;\;\;\;\;\;\;\;\;\;\;\;\left. +\left( \frac{i}{%
\left( 4\pi \right) ^{2}}\right) \left[ 2m^{2}-\left( k_{3}-k_{2}\right) ^{2}%
\right] \left( 2\xi _{01}\right) \right\}  \nonumber \\
&&-\frac{\varepsilon _{\mu \alpha \nu \lambda }}{4}\left( k_{3}-k_{1}\right)
_{\alpha }\left\{ I_{\log }\left( m^{2}\right) -\left( \frac{i}{\left( 4\pi
\right) ^{2}}\right) Z_{0}\left( \left( k_{1}-k_{3}\right) ^{2};m^{2}\right)
\right.  \nonumber \\
&&\;\;\;\;\;\;\;\;\;\;\;\;\;\;\;\;\;\;\;\;\;\;\;\;\left. +\left( \frac{i}{%
\left( 4\pi \right) ^{2}}\right) \left[ 2m^{2}-\left( k_{3}-k_{2}\right) ^{2}%
\right] \left( 2\xi _{10}\right) \right\}  \nonumber \\
&&-\frac{\varepsilon _{\mu \alpha \nu \lambda }}{4}\left[ \left(
k_{1}+k_{2}\right) _{\beta }+\left( k_{3}+k_{1}\right) _{\beta }\right]
\Delta _{\alpha \beta }  \nonumber \\
\bullet F_{\lambda \mu \nu } &=&\left( \frac{i}{\left( 4\pi \right) ^{2}}%
\right) \left( k_{3}-k_{1}\right) _{\xi }\left( k_{2}-k_{1}\right) _{\beta
}\left\{ \varepsilon _{\nu \beta \lambda \xi }[\left( k_{2}-k_{1}\right)
_{\mu }\left( \xi _{02}+\xi _{11}-\xi _{01}\right) \right. \\
&&\hspace{1in}\hspace{1in}\hspace{0.4in}+\left( k_{3}-k_{1}\right) _{\mu
}\left( \xi _{20}+\xi _{11}-\xi _{10}\right) ]  \nonumber \\
&&\hspace{1in}\hspace{1in}\hspace{0.1in}+\varepsilon _{\mu \beta \lambda \xi
}[\left( k_{2}-k_{1}\right) _{\nu }\left( \xi _{02}+\xi _{11}-\xi
_{01}\right)  \nonumber \\
&&\hspace{1in}\hspace{1in}\hspace{0.4in}+\left( k_{3}-k_{1}\right) _{\nu
}\left( \xi _{20}+\xi _{11}-\xi _{10}\right) ]  \nonumber \\
&&\hspace{1in}\hspace{1in}\hspace{0.1in}+\varepsilon _{\mu \beta \nu \xi
}[\left( k_{3}-k_{1}\right) _{\lambda }\left( \xi _{11}-\xi _{20}+\xi
_{10}\right)  \nonumber \\
&&\hspace{1in}\hspace{1in}\hspace{0.4in}+\left. \left( k_{2}-k_{1}\right)
_{\lambda }\left( \xi _{02}-\xi _{11}-\xi _{01}\right) ]\right\}  \nonumber \\
&&-\frac{\varepsilon _{\mu \nu \lambda \xi }}{4}\left\{ \left( \left(
k_{3}-k_{1}\right) _{\xi }+\left( k_{2}-k_{1}\right) _{\xi }\right) \varphi
_{0}\right\}  \nonumber \\
&&+\varepsilon _{\nu \beta \lambda \sigma }\left( k_{2}-k_{3}\right) _{\beta
}\frac{\Delta _{\mu \sigma }}{4}+\varepsilon _{\mu \beta \lambda \sigma
}\left( k_{2}-k_{3}\right) _{\beta }\frac{\Delta _{\nu \sigma }}{4}+  \nonumber
\\
&&+\varepsilon _{\mu \sigma \nu \beta }\left[ \left( k_{2}-k_{1}\right)
_{\beta }+\left( k_{3}-k_{1}\right) _{\beta }\right] \frac{\Delta _{\lambda
\sigma }}{4}  \nonumber
\end{eqnarray}
with 
\begin{equation}
\varphi _{0}=I_{\log }\left( m^{2}\right) +\left( \frac{i}{\left( 4\pi
\right) ^{2}}\right) \left\{ -Z_{0}\left( \left( k_{2}-k_{3}\right)
^{2};m^{2}\right) +2\left[ \frac{1}{2}+m^{2}\xi _{00}\right] -\left(
k_{3}-k_{1}\right) ^{2}\xi _{10}-\left( k_{2}-k_{1}\right) ^{2}\xi
_{01}\right\} .  \nonumber
\end{equation}
The functions $\xi _{mn}$ are related to the finite content of three point functions and are 
defined in appendix A. The complete solution to $AVV$ can be written in the following form 
\begin{eqnarray}
\frac{T_{\lambda \mu \nu }^{AVV}}{-4i} &=&\left( \frac{i}{\left( 4\pi
\right) ^{2}}\right) \left( k_{3}-k_{1}\right) _{\xi }\left(
k_{2}-k_{1}\right) _{\beta }\left\{ \varepsilon _{\nu \lambda \beta \xi
}[\left( k_{3}-k_{1}\right) _{\mu }\left( \xi _{20}+\xi _{11}-\xi
_{10}\right) \right.  \\
&&\hspace{1in}\hspace{1in}\hspace{0.4in}\;+\left( k_{2}-k_{1}\right) _{\mu
}\left( \xi _{11}+\xi _{02}-\xi _{01}\right) ]  \nonumber \\
&&\hspace{1in}\hspace{1in}\;\;+\varepsilon _{\mu \lambda \beta \xi }[\left(
k_{3}-k_{1}\right) _{\nu }\left( \xi _{11}+\xi _{20}-\xi _{10}\right)  
\nonumber \\
&&\hspace{1in}\hspace{1in}\hspace{0.4in}\;+\left( k_{2}-k_{1}\right) _{\nu
}\left( \xi _{02}+\xi _{11}-\xi _{01}\right) ]  \nonumber \\
&&\hspace{1in}\hspace{1in}\;\;+\varepsilon _{\mu \nu \beta \xi }[\left(
k_{3}-k_{1}\right) _{\lambda }\left( \xi _{11}-\xi _{20}+\xi _{10}\right)  
\nonumber \\
&&\hspace{1in}\hspace{1in}\hspace{0.4in}\;+\left. \left( k_{2}-k_{1}\right)
_{\lambda }\left( \xi _{02}-\xi _{01}-\xi _{11}\right) ]\right\}   \nonumber \\
&&+\left( \frac{i}{\left( 4\pi \right) ^{2}}\right) \frac{\varepsilon _{\mu
\nu \lambda \beta }}{4}\left( k_{3}-k_{1}\right) _{\beta }\left\{
Z_{0}\left( \left( k_{1}-k_{3}\right) ^{2};m^{2}\right) -Z_{0}\left( \left(
k_{2}-k_{3}\right) ^{2};m^{2}\right) \right.   \nonumber \\
&&\hspace{1in}\hspace{1in}+\left[ 2\left( k_{3}-k_{2}\right) ^{2}-\left(
k_{1}-k_{3}\right) ^{2}\right] \xi _{10}+  \nonumber \\
&&\hspace{1in}\hspace{1in}\left. +\;\left( k_{1}-k_{2}\right) ^{2}\xi _{01}+%
\left[ 1-2m^{2}\xi _{00}\right] \right\}   \nonumber \\
&&+\left( \frac{i}{\left( 4\pi \right) ^{2}}\right) \frac{\varepsilon _{\mu
\nu \lambda \beta }}{4}\left( k_{2}-k_{1}\right) _{\beta }\left\{
Z_{0}\left( \left( k_{1}-k_{2}\right) ^{2};m^{2}\right) -Z_{0}\left( \left(
k_{2}-k_{3}\right) ^{2};m^{2}\right) \right.   \nonumber \\
&&\hspace{1in}\hspace{1in}+\left[ 2\left( k_{3}-k_{2}\right) ^{2}-\left(
k_{1}-k_{2}\right) ^{2}\right] \xi _{01}  \nonumber \\
&&\hspace{1in}\hspace{1in}\left. +\;\left( k_{3}-k_{1}\right) ^{2}\xi _{10}+%
\left[ 1-2m^{2}\xi _{00}\right] \right\}   \nonumber \\
&&-\frac{\varepsilon _{\mu \nu \beta \sigma }}{4}\left[ \left(
k_{2}-k_{1}\right) _{\beta }+\left( k_{3}-k_{1}\right) _{\beta }\right]
\Delta _{\lambda \sigma }  \nonumber \\
&&+\frac{\varepsilon _{\nu \lambda \beta \sigma }}{4}\left(
k_{2}-k_{3}\right) _{\beta }\Delta _{\mu \sigma }+\frac{\varepsilon _{\mu
\lambda \beta \sigma }}{4}\left( k_{2}-k_{3}\right) _{\beta }\Delta _{\nu
\sigma }  \nonumber \\
&&-\frac{\varepsilon _{\mu \nu \lambda \alpha }}{4}\left[ \left(
k_{1}+k_{2}\right) _{\beta }+\left( k_{3}+k_{1}\right) _{\beta }\right]
\Delta _{\alpha \beta }.  \nonumber
\end{eqnarray}
It is important to emphasize that {\it this result is the most general one for the  $AVV$ 
amplitude. All possible choices for specific routing or shifts on loop momentum are 
automatically included. In addition this result can be mapped on those produced by specific 
regularization once the divergent
objects are maintained intact}. They are contained in the object 
$\Delta_{\alpha \beta }$.

In this result it is important to note the presence of a potentially ambiguous term, the last 
one in the above expression. This is due to the fact that the combinations of the $k_1$, $k_2$ 
and $k_3$ involved are not restricted to differences between them. As in the previously 
analyzed $\Pi _0^{\mu \nu }$ amplitude this term is a coefficient of the piece $\Delta$. 
In addition we can note that there is another term in the expression where the object 
$\Delta$ is present but with non ambiguous coefficients. Before more detailed analysis 
let us now to proceed in order to obtain the corresponding result for the Chern-Simons 
contribution. As an initial step in this direction we take the choices: 
$k_1=0$, $k_2=k_3=p$ in order to reproduce the kinematical situation corresponding to the 
first contribution for $\Pi^{\mu\nu\lambda}(p)$ in eq.(20). After 
this, in the 
obtained expression for $T^{AVV}_{\mu\nu\lambda}$, we make $k_1=p$, $k_2=k_3=0$ 
and 
interchange $\mu \leftrightarrow \nu$ to obtain the second contribution to 
eq.(20). Adding the results so obtained we get
\begin{eqnarray}
\Pi _{\lambda \mu \nu }\left( p\right)  &=&2i\left\{ \varepsilon _{\mu \nu
\beta \sigma }p^{\beta }\Delta _{\lambda \sigma }-\varepsilon _{\mu \nu
\lambda \beta }p^{\sigma }\Delta _{\beta \sigma }\right\}  \\
&&+\left( \frac{1}{4\pi ^{2}}\right) \varepsilon _{\mu \nu \lambda \beta
}p^{\beta }\left\{ Z_{0}\left( p^{2};m^{2}\right) -p^{2}\left( \xi _{01}+\xi
_{10}\right) +\left[ 1-2m^{2}\left( \xi _{00}\right) \right] \right\} . 
\nonumber
\end{eqnarray}
The contribution to the Chern-Simons term is extracted from the above expression 
taking $p^2=0$. Using the properties $Z_0(p^2=0)= 0$ and $\xi _{00}(p^2=0) = \frac{-1}{2m^2}$, 
the result is:
\begin{equation}
\Pi _{\lambda \mu \nu }\left( p\right) =\left( \frac{1}{2\pi ^{2}}\right)
\varepsilon _{\mu \nu \lambda \beta }p^{\beta }+2i\left\{ \varepsilon _{\mu
\nu \beta \sigma }p^{\beta }\Delta _{\lambda \sigma }-\varepsilon _{\mu \nu
\lambda \beta }p^{\sigma }\Delta _{\beta \sigma }\right\}   \nonumber
\end{equation}
As previously announced, from our result its easy to obtain the corresponding 
result of Jackiw 
and Kosteleck\'y. This result emerges immediately if we identify $-\Delta _{\xi \beta }$ with 
the surface term evaluated explicitly in ref. \cite{7}
\begin{eqnarray}
\Delta S_{\mu \nu } &=&\int_{\Lambda }\frac{d^{4}k}{\left( 2\pi \right) ^{4}}%
\frac{\partial }{\partial k_{\nu }}\left( \frac{k_{\mu }}{\left(
k^{2}-m^{2}\right) ^{2}}\right) =\int_{\Lambda }\frac{d^{4}k}{\left( 2\pi
\right) ^{4}}\frac{-4k_{\mu }k_{\nu }}{\left( k^{2}-m^{2}\right) ^{3}}%
+\int_{\Lambda }\frac{d^{4}k}{\left( 2\pi \right) ^{4}}\frac{g_{\mu \nu }}{%
\left( k^{2}-m^{2}\right) ^{2}} \\
&=&\left( \frac{i}{\left( 4\pi \right) ^{2}}\right) \left( \frac{1}{2}%
\right) g_{\mu \nu }  \nonumber
\end{eqnarray}
which leads us to
\begin{equation}
\Pi _{\mu \nu \lambda }(p)=\left( \frac 3{8\pi ^2}\right) \varepsilon _{\mu
\nu \lambda \beta }p^\beta .
\end{equation}
As should require to demonstrate the correctness of our approach.

If we stop the investigation at this point our calculations would represent 
one point of view for the problem identical to that adopted by Jackiw and 
Kosteleck\'y. Our contribution for this discussion would be reduced to just 
a more explicitly evaluation of the involved mathematical structures. It 
is time to add some new considerations for the discussions to deserve some attention 
of the reader. 

Let us then turn the attention back to our result but now following our arguments. 
The first aspect is about the value of $\Delta _{\xi \beta}$. Our discussion in 
the previous section taught us that a nonvanishing value for $\Delta _{\xi \beta}$ 
will not only render the Q.E.D. vacuum polarization tensor ambiguous but also will 
necessarily violate its gauge invariance. Therefore in order to maintain U(1) gauge 
invariance 
of $\Pi _0^{\mu \nu }$ we have to choose $\Delta _{\mu \nu }=0$ (which, as we have 
said, 
is immediately obtained in D.R.). Consequently the value for the Chern-Simons 
term, eq.(34), 
is given by  
\begin{equation}
\Pi _{\mu \nu \lambda }(p)=\left( \frac 1{2\pi ^2}\right) \varepsilon _{\mu
\nu \lambda \beta }p^\beta .
\end{equation}

The above results constitutes now, in fact, an alternative value for the 
Chern-Simons term once it is different from those presented by other authors 
in the literature. It is important to stress that to arrive at this value we 
have only chosen that the calculational method used to evaluate the 
divergent amplitudes \underline{exhibits} consistency in a larger context 
of the perturbative calculations. We can say that the so obtained result is 
dictated by the consistency requirement of the pure QED. In other words, 
the properties adopted to purely divergent integrals are necessary conditions 
to give meaningless to QED in the perturbative approach. Any regularization 
scheme, or equivalent philosophy that do not furnishes $\Box =\nabla =\Delta =0$ 
cannot be accepted for the treatment of the QED (in particular) divergences. 
In the D.R. technique these relations are automatically incorporated and this 
is exactly the reason for the success of this method in what concern to the 
ambiguity elimination and symmetry relations (Ward Identities) maintenance. 
In other QED consistent techniques like Pauli-Villars \cite{21} recipe, in the 
last instance, all we require to the coefficients of the superposition that 
characterizes the method, is that $\Box =\nabla =\Delta =0$, which can be 
easily verified \cite{21}. The relevant question, in light of these statements, 
is: can we accept a result produced by a strategy to handle divergences in 
perturbative calculations of QFT, for a particular problem, if the adopted 
strategy cannot lead us to a QED consistent results? If the answer is no, 
then the value in the eq.(37) represents a nonzero value for the Chern-Simons 
term free from the ambiguities related to arbitrary choices of the internal 
lines momenta in the loops. There is a very attractive aspect involved once 
we need not to admit the lost of the translational invariance in 
perturbative calculations, the most basic one implemented in a QFT. In spite 
of the preceding argumentation is bounded to QED, which is very strong by itself, we 
believe that these aspects are deeply and directly related to our problem 
for many reasons. Firstly $\Pi^{\mu\nu}_0$ present in the full amplitude, 
eq.(2), is a QED amplitude requiring then an identical treatment. For the 
second we can call to the discussions an important memory reason, the 
extended Standard Model, which have generate the discussion about the 
Chern-Simons term radiatively induced, developed by Colladay and Kosteleck\'y, 
were constructed in light of the renormalizability and, therefore, all 
assumptions that spoils the QED renormalizability destroys also the 
fundamental basis of the discussion. At this point if we take seriously 
the last comment then this is not the whole history. In the ref.\cite{5} it 
was conjectured by Colladay and Kosteleck\'y that the imposition of the 
anomaly cancellation consistency requirement, an important ingredient of the 
Standard Model, should lead to a vanishing value for the Chern-Simons term. 
Once the result present in our calculations, eq.(32), for the mathematical 
structures involved is the most general one, we are ready to investigate also 
this aspect. 

Up to now we have not asked what the U(1) gauge invariance of the $AVV$ amplitude can 
reveal to us? In different words; U(1) gauge invariance was assumed as an essential ingredient 
and is crucial in connection with the renormalizability by anomaly 
cancellation in the theory that has generated the amplitude $\Pi ^{\mu \nu }$. Then it seems not to 
make sense to attribute any significance to the result of a particular calculation that lost 
the initial ingredient, used as a guide to try to find a candidate to Lorentz invariance 
breaking phenomena. In summary: does the $AVV$ amplitude remain U(1) gauge invariant after the 
calculations? To adequately answer we need to verify its corresponding Ward identities. 
This is therefore what we will consider in the next section.

\section{Ward Identities and Anomalies}

In order to answer our present question we have to explicitly verify the Ward identities 
associated with $AVV$. This is obtained by contracting the expression (32) with the 
corresponding external momenta where from now on we adopt the definitions 
$k_3-k_1=q$, $k_1-k_2=p$, $k_3-k_2=p+q$ once only physical momenta remains in the 
expression after taking $\Delta=0$.  The detailed explicit evaluation of such identities 
is straightforward \cite{17}, \cite{20} although some algebraic effort is involved. The basic 
ingredients for this calculation involves properties of the functions $\xi _{mn}$ and $Z_k$. 
They are
\begin{eqnarray}
\bullet q^{2}\left( \xi _{11}\right) -\left( p\cdot q\right) \left( \xi
_{02}\right)  &=&\frac{1}{2}\left\{ \frac{-1}{2}Z_{0}\left(
(p+q)^{2};m^{2}\right) +\frac{1}{2}Z_{0}\left( p^{2};m^{2}\right)
+q^{2}\left( \xi _{01}\right) \right\}  \\
\bullet q^{2}\left( \xi _{20}\right) -\left( p\cdot q\right) \left( \xi
_{11}\right)  &=&\frac{1}{2}\left\{ -\left[ \frac{1}{2}+m^{2}\xi _{00}\right]
+\frac{p^{2}}{2}\left( \xi _{01}\right) +\frac{3q^{2}}{2}\left( \xi
_{10}\right) \right\}  \\
\bullet p^{2}\left( \xi _{02}\right) -\left( p\cdot q\right) \left( \xi
_{11}\right)  &=&\frac{1}{2}\left\{ -\left[ \frac{1}{2}+m^{2}\xi _{00}\right]
+\frac{q^{2}}{2}\left( \xi _{10}\right) +\frac{3p^{2}}{2}\left( \xi
_{01}\right) \right\}  \\
\bullet p^{2}\left( \xi _{11}\right) -\left( p\cdot q\right) \left( \xi
_{20}\right)  &=&\frac{1}{2}\left\{ -\frac{1}{2}Z_{0}\left(
(p+q)^{2};m^{2}\right) +\frac{1}{2}Z_{0}\left( q^{2};m^{2}\right) +p^{2}\xi
_{10}\right\}    \\
\bullet q^{2}\left( \xi _{10}\right) -\left( p\cdot q\right) \left( \xi
_{01}\right)  &=&\frac{1}{2}\left\{ -Z_{0}\left( \left( p+q\right)
^{2};m^{2}\right) +Z_{0}\left( p^{2};m^{2}\right) +q^{2}\left( \xi
_{00}\right) \right\}    \\
\bullet p^{2}\left( \xi _{01}\right) -\left( p\cdot q\right) \left( \xi
_{10}\right)  &=&\frac{1}{2}\left\{ -Z_{0}\left( \left( p+q\right)
^{2};m^{2}\right) +Z_{0}\left( q^{2};m^{2}\right) +p^{2}\left( \xi
_{00}\right) \right\} .
\end{eqnarray}
The result so obtained is:
\begin{eqnarray}
\bullet p^\nu T_{\lambda \mu \nu }^{AVV} &=&-\left(
\frac {1}{8\pi ^2}\right) \varepsilon _{\mu \nu \lambda \beta }p^\beta q^\nu  \\
\bullet q^\mu T_{\lambda \mu \nu }^{AVV} &=&-\left(
\frac 1{8\pi ^2}\right) \varepsilon _{\mu \nu \lambda \beta}p^\beta q^\mu  \\
\bullet \left( p+q\right)^\lambda T_{\lambda \mu \nu }^{AVV}&=&-2m\left\{T_{\mu \nu}^{PVV}\right\}, 
\end{eqnarray}
where 
\begin{equation}
T_{\mu \nu }^{PVV}=\left( \frac{-1}{4\pi ^2}\right) m \varepsilon _{ \mu \nu \lambda \beta } 
p^\beta q^\lambda \xi _{00},
\end{equation}
so the axial vector identity is satisfied and the two gauge identities violated.

These expressions do not constitute a surprising fact, once its the well known phenomena 
of anomaly involved in the pion decay \cite{16}. We only call attention to the way that 
they are obtained; completely off the mass shell, without the use of explicit regularizations 
and using a procedure that treats all amplitudes in all theories and models according to the 
same point of view.

The main aspect involved in our present discussion is the fact that we have lost the U(1) gauge 
invariance of the amplitude in the calculation. To attribute physical significance, in connection to 
the pion decay phenomenology  or Sutherland-Veltman paradox, a redefinition 
needs to be given by the inclusion of an anomalous term that allows us to recover the vector 
Ward identities \cite{16}
corresponding to our required U(1) gauge invariance. This could be achieved by the substitution 
\begin{equation}
\left(T_{\lambda \mu \nu }^{AVV}(p,q) \right)_{phys} =T_{\lambda \mu \nu }^{AVV}(p,q) 
-T_{\lambda \mu \nu }^{AVV}\left( 0\right) .
\end{equation}
Here $\left(T_{\lambda \mu \nu }^{AVV} \right)_{phys}$ is a redefinition of the amplitude 
and $T^{\lambda \mu \nu }_{AVV}(0) $ is its value at $p^2 = q^2 = 0$, namely 
\begin{equation}
\left(T_{\lambda \mu \nu }^{AVV} \right)(0) =\left( \frac 1{8\pi ^2}\right)
\varepsilon _{\mu \nu \lambda \beta }\left[ q^\beta - p^\beta \right] .
\end{equation}
After these considerations, the Ward identities will become: 
\begin{eqnarray}
\bullet p^\nu \left(T_{\lambda \mu \nu }^{AVV} \right)_{phy} &=&0 \\
\bullet q^\mu \left(T_{\lambda \mu \nu }^{AVV} \right)_{phy} &=&0 \\
\bullet \left( p+q \right) ^\lambda \left(T_{\lambda \mu \nu }^{AVV} \right)_{phy}
&=&-2m\left\{ T_{\mu \nu }^{PVV}\right\} -\left( \frac 1{4\pi ^2}\right)
\varepsilon _{\mu \nu \lambda \beta }\left[ p^\lambda q^\beta \right] .
\end{eqnarray}
The modified $AVV$ amplitude by the inclusion of the anomalous term is now in agreement 
with the phenomenology, so the right hand side of eq.(52), predicts the pion decay.
This modified $AVV$ amplitude is used to construct the renormalizable standard model by 
anomaly cancellation. 

After all these calculations a question is in order: in what sense $AVV$ triangle anomaly affect the 
Chern-Simons term? From the physical point of view a QFT in the perturbative approach is 
in last instance, a collection of basic amplitudes. These amplitudes are nothing more than 
mathematical structures (Green's functions) connected to the phenomenology by the insertion 
of the external lines which characterizes the physical processes. In a theory like the Standard 
Model different physical processes may be described perturbatively making use of the same intermediate 
Green's functions. The value attributed to the basic mathematical structures cannot be associated to the 
particular physical processes involved but needs to have the same value 
in all places of occurrence. The amplitude $\Pi_b^{\mu\nu}$, responsible for the 
Chern-Simons term, from the phenomenological point of view, in principle, there is nothing to 
do with anomalies in Ward Identities. In fact, only the value at the zero axial vertex momenta 
is required and on this kinematic point there is no violation in Ward Identities. When the 
anomalous term is added to the $AVV$ amplitude the $U(1)$ gauge invariance is present in all 
kinematic point that is the point of view adopted in the construction of the Standard Model 
by anomaly cancellation. If we follow this argumentation we must to import, when the 
$\Pi_b^{\mu\nu}$ calculation is in order, the anomalous term. \footnote{Note that for the 
QED extended implications, strictly speaking, this a choice and not a requirement. If 
however we take the QED extended as embedding in a more general theory then our specific 
choice became a requirement.}

We have finally arrived at the position where the answer to the question involved in this 
section can be furnished: the redefinition imposed by gauge invariance in the $AVV$ 
amplitude generates a subtraction that cancels exactly the remaining contribution eq.(37) to 
the value for the Chern-Simons term as previously conjectured in the literature \cite{5} 

\section{Summary and Discussions}

In this work we studied an extended version of Q.E.D. by the addition of an axial-vector 
term in Lagrangian 
$(\bar{\Psi}\nA{b} \gamma _5\Psi )$ in order to exploit the possibility of Lorentz and CPT 
symmetry violation induced by radiative corrections (to one-loop order). The relevant 
quantity to analyze is the vacuum polarization tensor $\Pi _{\mu \nu }$ which could be 
decomposed in three parts: one is structurally identical to the pure Q.E.D. vacuum 
polarization tensor $\Pi_0^{\mu \nu }$ and the others are  $\Pi _b^{\mu \nu }$ and 
$\Pi _{bb}^{\mu \nu }$ linear and quadratic in $b_\mu $ respectively. $\Pi _b^{\mu \nu }$ 
has the same mathematical form (closely related) as the famous $AVV$ triangle and 
$\Pi _{bb}^{\mu \nu }$ does not contribute \cite{7}.

Throughout the calculations we have not assumed any specific regularization scheme 
(or something equivalent) but instead we followed a simple strategy in dealing with 
the divergences: the integrals are assumed to be implicitly regulated by a function which 
is even in the integration variable and possesses a well defined connection limit. Under 
this assumption we could, by means of algebraic manipulation in the integrand, separate 
the finite content from the divergences. The latter were then displayed in terms of 
$I_{log}(m^2),I_{quad}(m^2)$ and differences between divergent integrals of same degree 
of divergences $\Box ,\nabla$ and $\Delta $ without any further assumption upon them. At 
this point we could readily map our procedure on the conventional ones (Dimensional 
Regularization, covariant Paulli-Villars, ...). This could be done by expressing 
$I_{log}(m^2)$ and $I_{quad}(m^2)$ according to the rules of the regularization scheme 
adopted and explicitly evaluating the differences $\Box ,\nabla$ and  $\Delta $. 
In doing so we can distinguish two classes of regularizations depending on the value 
of those differences being zero or not. For instance in D.R. it can be shown that 
$\Box ,\nabla$ and $\Delta $ vanish. In schemes like that adopted in the ref.[6] these  
differences can be identified as their surface terms.

Having set the grounds for our analysis, in order to verify if an induced Chern-Simons term 
could be generated radiatively we proceeded to a consistent treatment for the vacuum 
polarization tensor. To start with, we choose to demand the U(1) gauge invariance of 
$\Pi _0^{\mu \nu }$ (pure Q.E.D.). This in turn implied that $\Box =\nabla =\Delta =0$. Now 
it is important to realize that these objects appear in the expression for $\Pi _{\mu \nu }$ 
with coefficients which are {\it not} ambiguous i.e. dependent on the combination $k_1-k_2$. 
However two of them $\Box $ and $\Delta $ do have ambiguous coefficients ($k_1$ and $k_2$ 
in other combinations than differences). Therefore there is no possibility to find 
a particular routing of the internal momenta so as to maintain gauge invariance. Thus 
gauge invariance itself eliminates the ambiguous terms. In this sense the consistency 
conditions $(\Box =\nabla =\Delta =0)$ produce the same result as in DR. 
With this result in mind we proceeded to calculate the second term $\Pi_b^{\mu \nu }$, 
which can be identified with a particular kinematical situation of the $AVV$ amplitude. 
The most general explicit form of this amplitude was obtained in such way that all 
possibilities are still present. From the result so obtained three clearly situations, 
corresponding to different values for the Chern-Simons term, can be identified.

{\bf i} \underline{Surface's terms evaluation}.

In our general result eq.(32) if we interpret the divergent content remaining, 
represented by the $\Delta_{\mu\nu}$ object, as a surface term
\begin{equation}
-\Delta_{\mu\nu}=S_{\mu\nu}=\left(\frac{i}{(4\pi )^2}\right)
\left(\frac{1}{2}\right)g_{\mu\nu}
\end{equation}
the value for the Chern-Simons term is given by
\begin{equation}
\Pi_{\mu \nu \lambda}(p)=\left(\frac{3}{8\pi^2}\right)
\varepsilon_{\mu\nu\lambda\beta}p^\beta
\end{equation}
which corresponds to that obtained by other authors. In particular, this is the point 
of view adopted for the problem in ref.\cite{1}. Our criticism in respect to this 
result resides on the fact that there is a hard price to pay in adopting this way. 
All perturbative divergent amplitudes, with divergence degree higher than logarithmic, 
are assumed as ambiguous quantities and the symmetry relations may be violated as we have 
showed analyzing the $\Pi_0^{\mu \nu }$ 
term for the full expression of the present problem.

{\bf ii} \underline{U(1) gauge invariance of the $\Pi_0^{\mu \nu }$}.

Guided by the consistency requirements of the calculational method used to evaluate 
divergent amplitudes in a more general sense we have learned that we need to choose 
$\Delta_{\mu\nu}=0$ (as well as $\Box_{\alpha\beta\mu\nu}=\nabla_{\mu\nu}=0$). With 
this interpretation all the results obtained by our calculational approach can be mapped 
in those corresponding to D.R. calculations in all places where this technique can be 
applied. In consequence the amplitudes 
will turn out free from the ambiguities associated to the arbitrary choices involved in 
the rotulation of the internal lines momenta in loops. In this context, the value for the 
Chern-Simons term so obtained is given by 
\begin{equation}
\Pi_{\mu \nu \lambda}(p)=\left(\frac{1}{2\pi^2}\right)
\varepsilon_{\mu\nu\lambda\beta}p^\beta ,
\end{equation}
which is, in consequence, ambiguities free. We can say that it is determined by the QED 
renormalizability consistency requirements in perturbative calculations.

{\bf iii} \underline{Anomaly cancellation implications}

In the expression obtained for $AVV$, once we eliminated the ambiguous term, we verified that 
the vector Ward identities were violated whereas the axial was obtained satisfied. 
{\it This is the same situation as we encounter in the pion decay where phenomenology 
tells us that an anomaly term must be added. This is done by choosing that U(1) gauge 
invariance is maintained. Thus we redefined $AVV$ by subtracting the anomalous term. 
Consequently we have a result that is gauge invariant for all momenta (not only at zero)}. 

In proceeding this way we adopted the same value for the mathematical structure involved 
in the 
present problem, the $AVV$ Green's function, as in the Standard Model construction by anomaly 
cancellation.

The anomalous term included exactly cancels the value obtained in $\Pi _{\mu \nu \lambda }(p)$ 
in eq.(37) which naturally establishes that $k_\mu $ vanishes according to the conjecture of 
the Colladay and Kosteleck\'{y} \cite{5},
\begin{equation}
\Pi_{\mu \nu \lambda}(p)=0.
\end{equation}
 
All we required to obtain our conclusion was $U(1)$ gauge invariance in addition to the 
procedure that we adopted to manipulate and compute the divergent quantities. In virtue 
of these conclusions we can give a safe answer to the question put on the title of this 
contribution: What is the role played by ambiguities in calculations of the radiatively 
induced Chern-Simons shift in extended Q.E.D.? \underline{ Gauge invariance leaves no room 
for ambiguities} in perturbative calculations. This conclusion, extracted from our analysis 
in the present problem, actually remains valid in all other contexts where perturbative 
calculations is applied.

This procedure can be used to treat any problem that involves divergences in perturbative 
calculations at any order in $\hbar $ (in Q.F.T.), for both abelian and non-abelian theories, 
as well as to calculate renormalization group coefficients \cite{22}. The results obtained this 
way are ambiguity free. This framework yields a clear formulation of anomalies. The Ward 
identities are automatically satisfied where they should. Applications in the context of 
renormalizable and non-renormalizable theories have been successfully affected. The main 
advantage resides in the simplicity and clarity once no explicit form of regularization 
is needed. In other words divergent integrals are {\it not} explicitly calculated at any 
step of the calculations. All we need to require are general properties and the so called
 consistency conditions. Consequently if a regularization satisfies the consistency conditions
 it is thus not necessary.

{\bf Acknowledgements}: We are indebted to M.C. Nemes and M. Sampaio for most fruitful 
discussions and to C.O. Gra\c ca for a careful reading of this manuscript.

\appendix

\section{ General Integrals for the Finite Content of One Loop Amplitudes}

\noindent {\bf The Functions $Z_k(p^2;m^2)$}

We define 
\begin{equation}
Z_k(p^2;m^2)=\int_0^1dz\,z^kln\,\left( {\frac{p^2z(1-z)-m^2}{-m^2}}\right), 
\end{equation}
where $k$ is an integer, $m$ is the mass parameter which appears in the propagators, 
$p$ is some external momentum.

\noindent {\bf The Functions $\xi _{nm}$}

When we consider one loop Feynman integrals associated to three point functions with two 
external momenta, the finite parts of the amplitudes are always related to the following 
general structures (same masses) 
\begin{equation}
\xi _{nm}(p,q)=\int_0^1\,dz\int_0^{1-z}\,dy{\frac{z^ny^m}{Q(y,z)}},
\end{equation}
where 
\begin{equation}
Q(y,z)=p^2y(1-y)+q^2z(1-z)-2(p\cdot q)yz-m^2.
\end{equation}

\section{ Divergent Integrals}

All the integrals below are divergent. We use the identity (8) to separate the divergent 
part of the finite part. Only the finite part of the integrals should depend on the external 
momenta. The remaining divergent integrals, now independent of the physical momenta, are 
organized in terms of a set of differences between divergent integrals eq.(9),(10),(11) and 
the basic divergent objects $I_{log}(m^2)$ and $I_{quad}(m^2)$. With this philosophy we show 
bellow some divergent integrals that are necessary to calculate the amplitude $\Pi _{\mu \nu }$.
Thus 

\begin{eqnarray}
&&\bullet \int_{\Lambda }\frac{d^{4}k}{(2\pi )^{4}}\frac{1}{%
[(k+k_{1})^{2}-m^{2}]}=I_{quad}(m^{2})+k_{1\mu }k_{1\nu }\Delta _{\mu \nu }\\
&&\bullet \int_{\Lambda }\frac{d^{4}k}{(2\pi )^{4}}\frac{1}{%
[(k+k_{1})^{2}-m^{2}][(k+k_{2})^{2}-m^{2}]}=I_{log}(m^{2})-\left( \frac{i}{%
(4\pi )^{2}}\right) Z_{0}(\left( k_{1}-k_{2}\right) ^{2};m^{2})\;\;\;\;\;\;\; \\
&&\bullet \int_{\Lambda }\frac{d^{4}k}{(2\pi )^{4}}\frac{k_{\mu }}{%
[(k+k_{1})^{2}-m^{2}][(k+k_{2})^{2}-m^{2}]}=-2\left( k_{1}+k_{2}\right)
_{\alpha }\int_{\Lambda }\frac{d^{4}k}{(2\pi )^{4}}\frac{k_{\mu }k_{\alpha }%
}{\left( k^{2}-m^{2}\right) ^{3}} \\
&&\;\;\;\;\;\;\;\;\;\;\;\;\;\;\;\;\;\;\;\;\;\;\;\;\;\;\;\;\;\;\;\;\;\;\;\;\;%
\;\;\;\;\;\;\;\;\;+(k_1+k_2)_\mu\left( \frac{i}{(4\pi )^{2}}\right)
Z_{1}(\left( k_{1}-k_{2}\right) ^{2};m^{2})  \nonumber \\
&&\bullet \int_{\Lambda }\frac{d^{4}k}{(2\pi )^{4}}\frac{k_{\mu }k_{\nu }}{%
[(k+k_{1})^{2}-m^{2}][(k+k_{2})^{2}-m^{2}]}=\int_{\Lambda }\frac{d^{4}k}{%
(2\pi )^{4}}\frac{k_{\mu }k_{\nu }}{\left( k^{2}-m^{2}\right) ^{2}}-\left(
k_{1}^{2}+k_{2}^{2}\right) \int_{\Lambda }\frac{d^{4}k}{(2\pi )^{4}}\frac{%
k_{\mu }k_{\nu }}{\left( k^{2}-m^{2}\right) ^{3}}  \nonumber \\
&&+\left( k_{1\alpha }k_{1\beta }+k_{2\alpha }k_{2\beta }+k_{1\alpha
}k_{2\beta }\right) \int_{\Lambda }\frac{d^{4}k}{(2\pi )^{4}}\frac{4k_{\mu
}k_{\nu }k_{\alpha }k_{\beta }}{\left( k^{2}-m^{2}\right) ^{4}}  \nonumber \\
&&+\left( \frac{i}{(4\pi )^{2}}\right) \left\{ -(k_{1}-k_{2})_{\mu
}(k_{1}-k_{2})_{\nu }Z_{2}(\left( k_{1}-k_{2}\right) ^{2};m^{2})\right.  
\nonumber \\
&&\;\;\;\;\;\;\;\;\;\;\;\;\;\;-\left( k_{1}-k_{2}\right) ^{2}g_{\mu \nu }
\left[ \frac{1}{4}Z_{0}(\left( k_{1}-k_{2}\right) ^{2};m^{2})-Z_{2}(\left(
k_{1}-k_{2}\right) ^{2};m^{2})\right]   \nonumber \\
&&\;\;\;\;\;\;\;\;\;\;\;\;\;\;+k_{1\mu }\left[ \left( k_{1}-k_{2}\right)
_{\nu }Z_{1}(\left( k_{1}-k_{2}\right) ^{2};m^{2})\right]   \nonumber \\
&&\;\;\;\;\;\;\;\;\;\;\;\;\;\;+k_{1\nu }\left[ \left( k_{1}-k_{2}\right)
_{\mu }Z_{1}(\left( k_{1}-k_{2}\right) ^{2};m^{2})\right]   \nonumber \\
&&\;\;\;\;\;\;\;\;\;\;\;\;\;\;-\left. k_{1\mu }k_{1\nu }Z_{0}(\left(
k_{1}-k_{2}\right) ^{2};m^{2})\right\} .  \nonumber
\end{eqnarray}
\begin{eqnarray}
&&\bullet \int_{\Lambda }\frac{d^{4}k}{(2\pi )^{4}}\frac{1}{%
(k^{2}-m^{2})[(p+k)^{2}-m^{2}][(q+k)^{2}-m^{2}]}=\left( \frac{i}{(4\pi )^{2}}%
\right) \xi _{00}(p,q) \\
&&\bullet \int_{\Lambda }\frac{d^{4}k}{(2\pi )^{4}}\frac{k_{\mu }}{%
(k^{2}-m^{2})[(p+k)^{2}-m^{2}][(q+k)^{2}-m^{2}]}=-\left( \frac{i}{(4\pi )^{2}%
}\right) \{q_{\mu }\xi _{10}(p,q)+p_{\mu }\xi _{01}(p,q)\}\;\;\;\;\;\;\;\; \\
&&\bullet \int_{\Lambda }\frac{d^{4}k}{(2\pi )^{4}}\frac{k_{\mu }k_{\nu }}{%
(k^{2}-m^{2})[(p+k)^{2}-m^{2}][(q+k)^{2}-m^{2}]}=\int_{\Lambda }\frac{d^{4}k%
}{(2\pi )^{4}}\frac{k_{\mu }k_{\nu }}{\left( k^{2}-m^{2}\right) ^{3}} \\
&&\;\;\;-\left( \frac{i}{(4\pi )^{2}}\right) \left\{ g_{\mu \nu }\left[ {%
\frac{1}{2}}Z_{0}((p-q)^{2};m^{2})-\left( {\frac{1}{2}}+m^{2}\xi
_{00}(p,q)\right) +{\frac{q^{2}}{2}}\xi _{10}(p,q)+{\frac{p^{2}}{2}}\xi
_{01}(p,q)\right] \right.   \nonumber \\
&&\;\;\;\;\;\;\;\;\;\;\;\;\;\;\;\;\;\;\;-\left. p_{\mu }p_{\nu }\xi
_{02}(p,q)-q_{\mu }q_{\nu }\xi _{20}(p,q)-(p_{\mu }q_{\nu }-p_{\nu }q_{\mu
})\xi _{11}(p,q)\right\} .  \nonumber
\end{eqnarray}


\begin{thebibliography}{99}
\bibitem{1} 
See, for example, S. Weinberg, ``The Quantum Theory of Fields", Volume II, Cambridge, 1996. 

\bibitem{2} 
P.H. Frampton, ``Gauge Fields Theories", Benjamin/Cummings, Menlo Park, California 1987.

\bibitem{3} 
R.M. Barnett {\it et al.}, Review of Particle Properties, Pys. Rev. D{\bf 54} (1996) 1;

L.K. Gibbons {\it et al.}, Fermilab-Pub-95/392-E (January 1996); B. Schwingenheuer {\it et al.},
Phys. Rev. Lett. {\bf 74} (1995) 4376;

R. Carosi  {\it et al.}, Phys. Rev. B{\bf 237} (1990) 303.

\bibitem{4} 
D.Colladay and V.A. Kosteleck\'{y}, Phys. Rev. D{\bf 55}, 6760 (1997).

\bibitem{5}
D.Colladay and V.A. Kosteleck\'{y}, Phys. Rev. D{\bf 58}, 116002 (1998).

\bibitem{6}
S. Coleman and S. Glashow, Phys. Rev. D{\bf 59}, 116008 (1999).

\bibitem{7}  
R. Jackiw and V. Alan Kosteleck\'{y}, Phys. Rev. Lett. {\bf 82}, 3572-3575 (1999).

\bibitem{8}
R. Jackiw and S. Templeton, Phys. Rev. D{\bf 23}, 2291 (1981);

J. Schonfeld, Nucl. Phys. B{\bf 185}, 157 (1981);

S. Deser, R. Jackiw and S. Templeton, Ann. Phys. (NY) {\bf 140}, 372 (1982).

\bibitem{9}
S. Carroll, G. Field and R. Jackiw, Phys. Rev. D{\bf 41}, 1231 (1990).

\bibitem{10}
M. Goldhaber and V. Trimble, J. Astrophys. Astr. {\bf 17}, 17 (1996);

S. Carroll and G. Field, Phys. Rev. Lett. {\bf 79}, 2394 (1997).

\bibitem{11} 
J.S. Chung and P. Oh, MIT-CTP-2809, HEP-TH/9812132.

\bibitem{12} 
W. F. Chen, HEP-TH/9903258.
 
\bibitem{13}
M. P\'{e}rez-Victoria, Phys. Rev. Lett. {\bf 83}, 2518 (1999).

\bibitem{14}
G.'t Hooft and M. Veltman, Nucl. Phys. B{\bf 44}, 189 (1972);

C.G. Bollini and J.J. Giambiagi, Phys Lett. B{\bf 40}, 566 (1972);

J.F. Ashmore, Nuovo Cimento Lett. {\bf 4}, 289 (1972).

\bibitem{15}
M. Chanowitz, M. Furman and I. Hinchliffe, Nucl. Phys. B{\bf 159}, 225 (1979);

P. Ramond, Field Theory: A modern Primer, Addisson-Wesley (1990). 

\bibitem{16} 
L.S. Gernstein and R. Jackiw, {\it Phys. Rev.} {\bf 181}, (1969) 5; 
See also J.S. Bell and R. Jackiw, {\it Nuovo Cimento} {\bf 60}, (1969) 47.

\bibitem{17}
O.A. Battistel, {\it PhD Thesis 1999}, Universidade Federal de Minas Gerais, Brazil.

\bibitem{18}  
See E.g. N.N. Bogoliubov and D.V. Shirkow, {\it Introduction to
the theory of Quantized Fields} Interscience Publ. Inc. 1959.

\bibitem{19}  
O.A. Battistel and M.C. Nemes, Phys. Rev. D{\bf 59} 055010 (1999).

\bibitem{20} 
G. Dallabona, {\it Master Thesis 1998}, Universidade Federal de Minas Gerais, Brazil.

\bibitem{21} 
O.A. Battistel, A.L. Mota and M.C. Nemes, Mod. Phys. Lett. A{\bf 13}, 1557 (1998).

\bibitem{22}  
A. Brizola, O.A. Battistel, M. Sampaio, M.C. Nemes, Mod. Phys. Lett. A{\bf 14}, 1509 (1999).

\end{thebibliography}
\end{document}